\documentclass[12pt]{iopart}
\usepackage{graphicx}
\expandafter\let\csname equation*\endcsname\relax
\expandafter\let\csname endequation*\endcsname\relax
\usepackage{amsmath}
\begin{document}
\title{NiSi$_2$ as seed layer for epitaxial growth of NiAl and Cr on Si(001)}

\author{M Ben Chroud$^{1,2}$, T-H T Tran$^{1,3}$, J Swerts$^1$, K Temst$^{2,1}$ and R Carpenter$^1$}

\address{$^1$ Imec, Kapeldreef 75, 3001 Leuven, Belgium}
\address{$^2$ Quantum Solid State Physics division, Department of Physics and Astronomy, KU Leuven, Celestijnenlaan 200D, 3001 Leuven, Belgium}
\address{$^3$ Surface and Interface Engineered Materials division, Department of Materials Engineering, KU Leuven, Kasteelpark Arenberg 44, 3001 Leuven, Belgium}
\ead{mohamed.benchroud@imec.be}

\begin{abstract}
In general, metal layers cannot be grown epitaxially on Si due to the tendency of metals to react and form a silicide. Even in case the metal layer has a matching lattice symmetry and atomic distance, the Si/metal interface is disturbed by the silicide thus preventing epitaxial growth. One exception is NiAl which is known to grow epitaxially when deposited on Si(001). During the growth, NiAl reacts with Si to form NiSi$_2$. In this work, epitaxial NiAl is grown and significant silicidation is observed in accordance with previous reports. However, the role that this silicide plays as a template for the epitaxial growth of NiAl has not been clear to this date. We hypothesize that NiSi$_2$ acts as a necessary seed layer between the Si substrate and the NiAl layer. Additionally, NiSi$_2$ can be used as a seed layer for the epitaxial growth of other metals besides NiAl. This was tested by growing NiSi$_2$ seperately and replacing the NiAl layer with Cr. Growing Cr directly on Si(001) produced a polycrystalline layer. When NiSi$_2$ was used as a seed layer, the Cr layer was found to be a single crystal with Si(001)//Cr(001) and Si(100)//Cr(100). NiSi was also tried as seed layer for Cr and was found to produce a polycrystalline Cr layer. Using NiSi$_2$ as a seed layer could enable the growth of various epitaxial materials for industrial semiconductor applications. 
\end{abstract}
\\
\section{Introduction}
Single crystal thin films often exhibit superior properties compared to their polycrystalline counterparts. One example is reduced resistivity in metal thin films \cite{milosevic19,benchroud24}. Low-resistive epitaxial materials have the potential to replace polycrystalline Cu as interconnect material in CMOS chips \cite{barmak20}. Cu is currently hitting its limits when scaling down interconnects since the resistivity rapidly increases at thicknesses below 10\,nm. In a single crystal film grain boundaries are minimized thus reducing the grain boundary contribution to the overall resistivity of the material \cite{mayadas70}. A second example is epitaxial BaTiO$_3$ which can be scaled down more effectively than polycrystalline BaTiO$_3$ while maintaining ferroelectricity \cite{kim12}. A third example comes from magnetoresistive random-access memory (MRAM) which relies on tunnel magnetoresistance (TMR) for reading of the memory. The TMR effect results from spin-dependent tunneling of current through a ferromagnet/barrier/ferromagnet multilayer stack e.g., Fe/MgO/Fe. TMR is strongly influenced by the MgO and Fe crystallinity with the highest TMR value being measured in an epitaxial layer stack \cite{scheike23}.
\\\\
A major roadblock for implementation of epitaxial materials in the semiconductor industry is the need for specialized substrates e.g., MgO and sapphire. Such substrates are costly and are hard to scale when compared to the industry standard 300\,mm Si wafer. One drawback of Si wafers is the difficulty of growing epitaxial metals since they readily react with the wafer to form a metal silicide. To overcome this challenge, a seed layer can be deposited on the Si wafer such that the desired epitaxial material can be grown on top. NiAl has been shown to be a suitable seed layer for layer stacks where ordinarily a MgO substrate would be used \cite{chen16,yakushiji19}. Epitaxial NiAl on Si has been used as a seed layer for a magnetic layer stack containing a Heusler alloy and Ag with the goal of improving the giant magnetoresistance (GMR) effect \cite{chen16}. GMR originates from spin-dependent electron scattering in a multilayer heterostructure consisting of alternating ferromagnetic and non-magnetic metallic layers. The magnetoresistance was found to be the same value compared to layer stacks grown on a MgO(001) substrate and outperformed its polycrystalline counterpart. Silicide formation did occur between the NiAl and Si which was identified to be NiSi$_2$ \cite{chen16}.
\\\\
NiSi$_2$ has a CaF$_2$-type crystal structure with a lattice parameter of $5.398\,\text{\AA}$ \cite{smeets09} which is only 0.6\% smaller than the diamond lattice of Si. As a consequence NiSi$_2$ can grown epitaxially on Si with a crystal orientation of NiSi$_2$(001)//Si(001) and NiSi$_2$(100)//Si(100). In the case of NiAl, NiSi$_2$ formation occurs as the NiAl reacts with the Si substrate. Epitaxial NiSi$_2$ can also be formed by reacting a Ni thin film with Si by annealing at 800$^\circ$C \cite{chiu81}. Annealing at lower temperatures, between $\sim$350$^\circ$C and 800$^\circ$C, results in NiSi formation \cite{dekeyser10,deschutter16A}. The aforementioned 800$^\circ$C annealing required for NiSi$_2$ can be reduced when the Ni layer is reduced to below 4\,nm. This was first reported on by \textit{Tung et al.} showing that epitaxial NiSi$_2$ forms by annealing 2\,nm of Ni at 450$^\circ$C \cite{tung83}. Later findings showed that an epitaxial nickel silicide forms at an annealing temperature of 300-400$^\circ$C and a Ni layer thickness below 3.7\,nm \cite{dekeyser10}. To this day, there is no proven explanation to why the NiSi$_2$ formation temperature drops for such thin Ni layers. One possible explanation originates from the, presumably, low interfacial energy of the epitaxial Si/NiSi$_2$ interface. This would favor formation of NiSi$_2$ over NiSi for thin Ni layers. For thicker layers, the interfacial energy has less of an influence as the bulk free energy is dominating.
\\\\
The purpose of this work is to elucidate the role NiSi$_2$ plays in the growth of epitaxial NiAl. The hypothesis is that NiSi$_2$ acts a bridging layer between the diamond lattice of Si and the bcc lattice of NiAl. In other words, NiSi$_2$ is not simply a byproduct formed during NiAl deposition but instead it is essential for epitaxial growth. To test this hypothesis, NiSi$_2$ was grown followed by a layer of Cr. Cr has the same lattice as NiAl and a nearly identical lattice parameter differing by less than 0.1\%. If NiSi$_2$ is indeed a bridging layer, then Cr should grow epitaxially. As a comparison, NiSi was also tried out as a seed layer as well as Cr deposited directly on Si.

\section{Experimental}
\begin{figure}
    \centering
    \includegraphics[width=10cm]{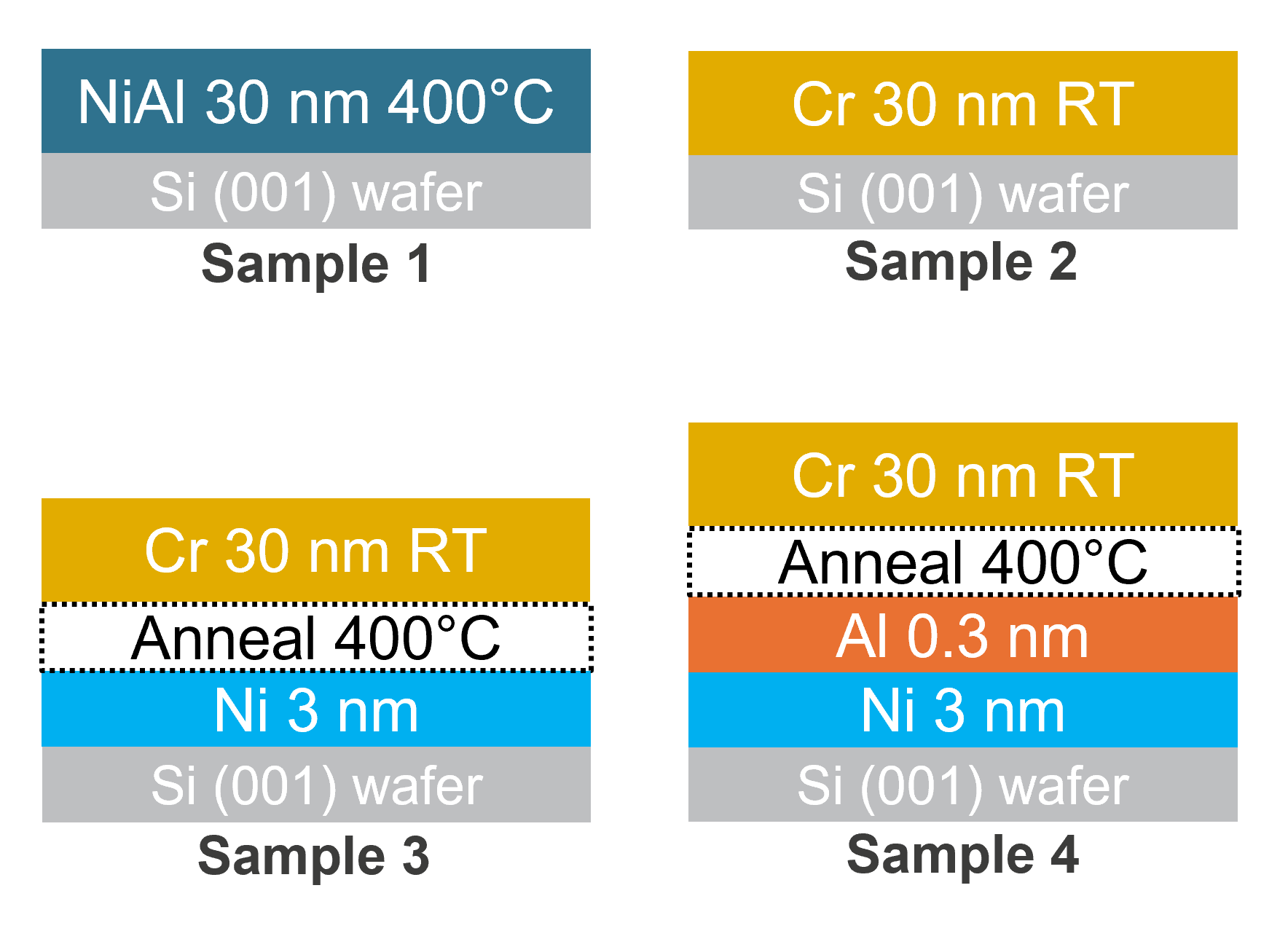}
    \caption{Deposition and anneal steps of the four samples.}
    \label{fig:overview}
\end{figure}

The starting substrate was a single crystal 300\,mm Si wafer with (001) orientation. The native surface oxide was removed by a dip in a 0.7\% HF solution for 60\,s. All deposition and annealing steps were done without braking the vacuum in a Canon Anelva EC7800 magnetron sputtering tool. The base pressure was in the order of $10^{-7}$\,Pa. All sputtering was carried out in DC mode using Ar gas at a working pressure in the order of 0.1\,Pa. Sample 1 had 30\,nm of NiAl, sputtered at a wafer temperature of 400$^\circ$C. This temperature has been shown before to produce an epitaxial NiAl layer \cite{chen16,yakushiji19}. The target consisted out of an NiAl alloy with a 1:1 atomic concentration. The other 3 samples had 30\,nm of Cr, deposited at room temperature (RT), with varying seed layers, as shown in figure \ref{fig:overview}. For sample 2, the Cr layer was grown directly on Si. Sample 3 had a seed layer of 3\,nm Ni annealed at 400$^\circ$C for 5\,min. Lastly, sample 4 was the same as sample 3 except for an additional Al layer of 0.3\,nm. Adding such a layer of Al to Ni has been shown to promote NiSi$_2$ growth for Ni thicknesses up to 6\,nm \cite{allenstein05,geenen18}.
\\
\\
The surface roughness was measured with tapping mode atomic force microscopy (AFM). The AFM images consisted of 512 lines which were scanned at a rate of 0.5 lines/s. Scan area sizes of 2x2\,\textmu m and 20x20\,\textmu m were used. X-ray diffraction (XRD) measurements were done on a Rigaku SmartLab®. The X-ray source was 9\,kW rotating anode with a 2-bounce Ge(220) monochromator resulting in pure Cu K$_{\alpha1}$ X-rays. The diffracted beam was measured by a 2D detector. Transmission electron microscopy (TEM) was used to assess thin film crystallinity and energy-dispersive X-ray spectroscopy (EDS) was used to measure atomic concentrations. Both TEM and EDS were done using a FEI Titan$^3$ at operating at 200\,kV. The TEM/EDS lamellae were cut by focused ion beam using a Helios 450.

\section{Results}
\begin{figure}
    \centering
    \includegraphics[width=10cm]{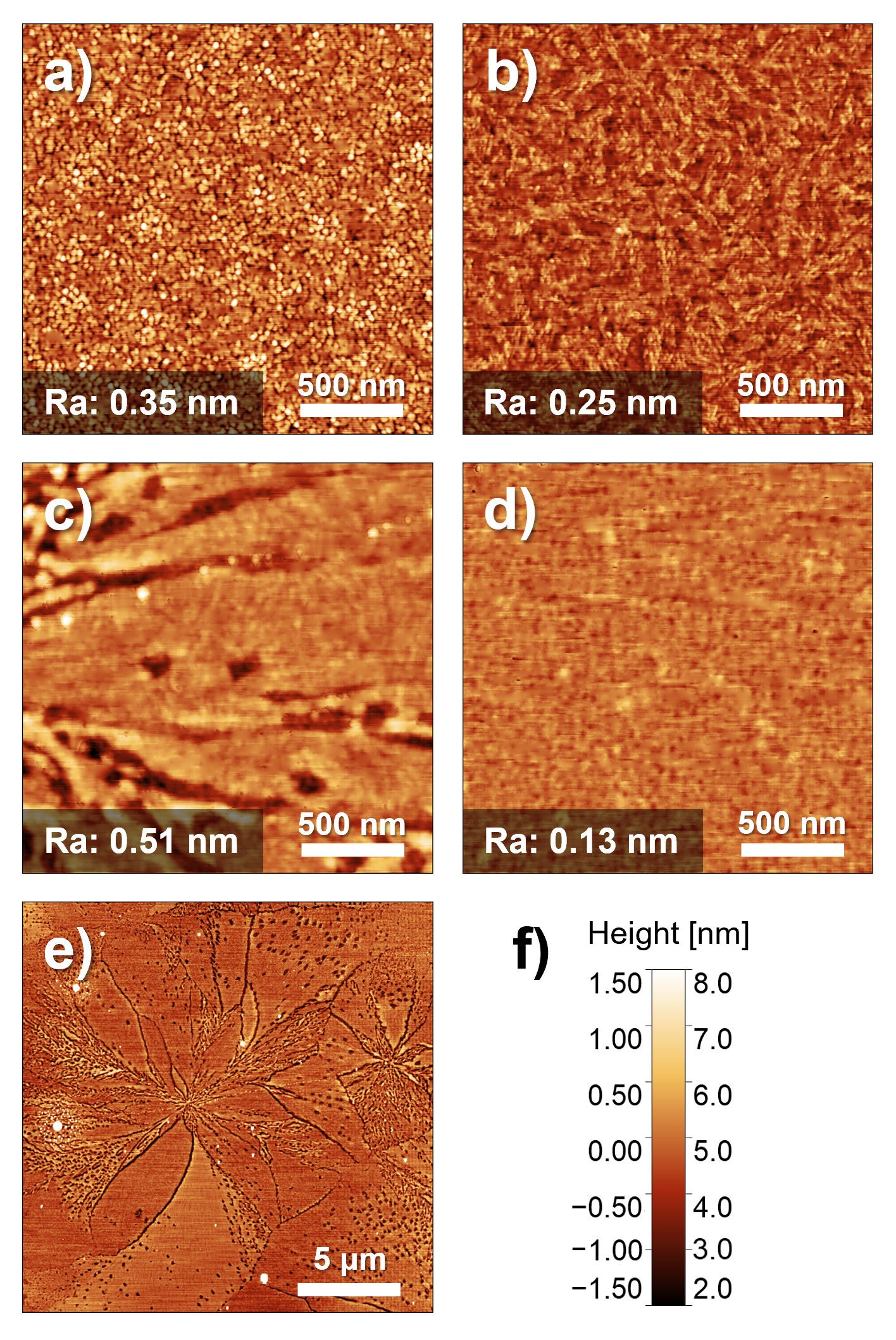}
    \caption{\textbf{a-d)} 2x2\,\textmu m-sized AFM scans of samples 1,2,3, and 4, respectively. \textbf{e)} 20x20\,\textmu m-sized scan of sample 3. \textbf{f)} height scale: The numbers on the left are for a),b), and d). The numbers on the right are for c) and e).}
    \label{fig:AFM}
\end{figure}

\begin{figure}
    \centering
    \includegraphics[width=10cm]{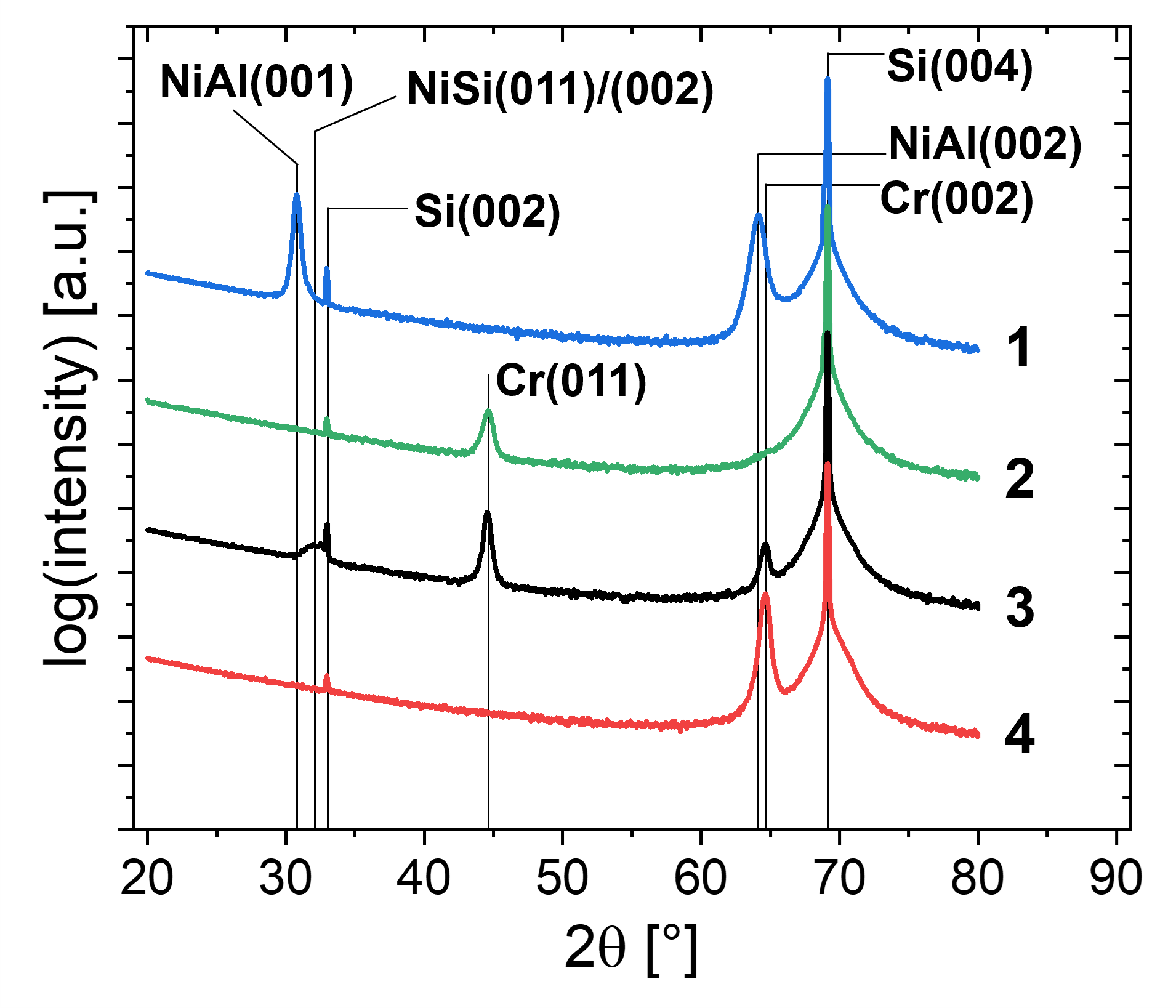}
    \caption{XRD 2$\theta$-$\theta$-scan of samples 1,2,3, and 4.}
    \label{fig:T2T}
\end{figure}

\begin{figure}
    \centering
    \includegraphics[width=10cm]{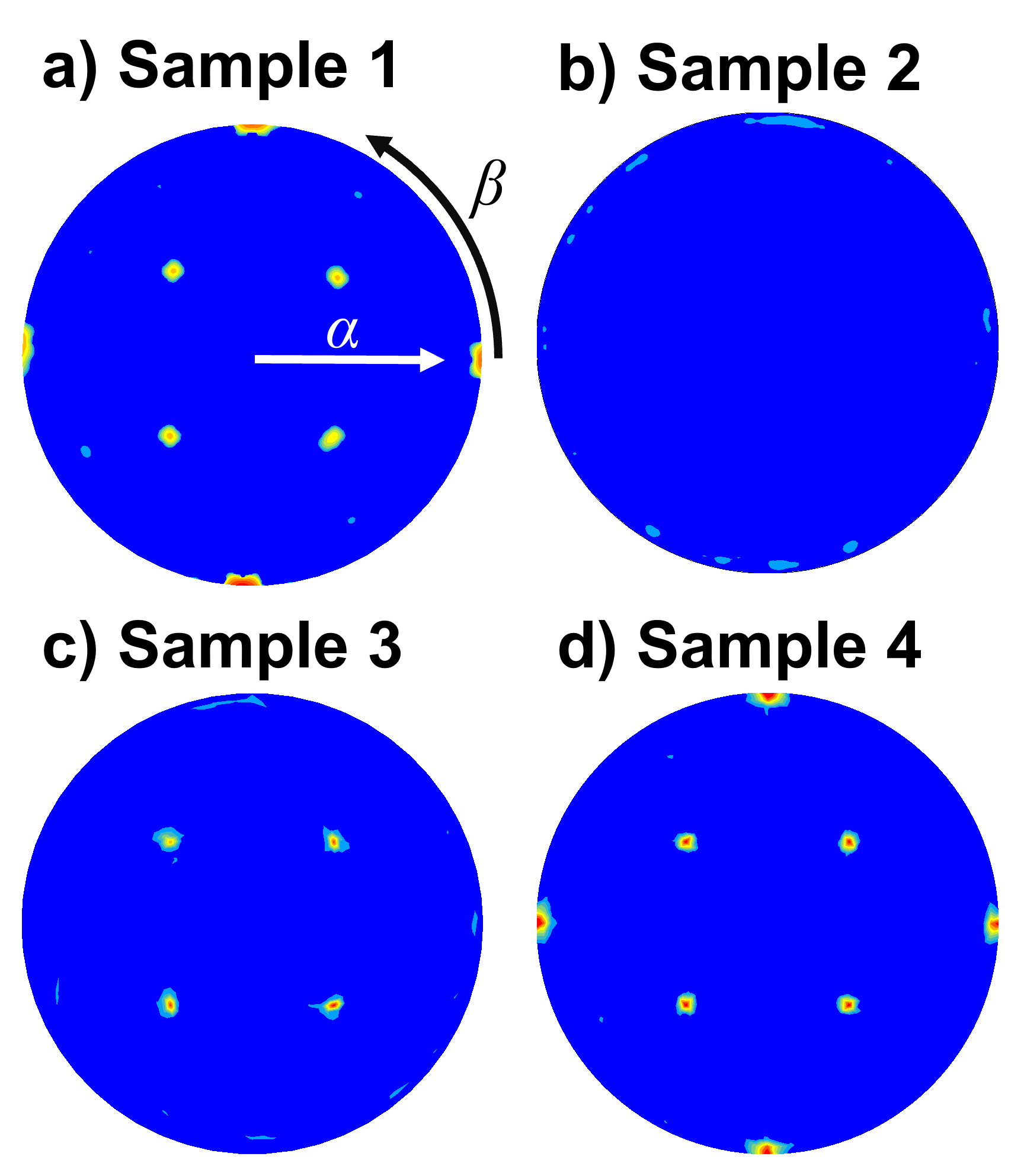}
    \caption{XRD pole figures of the four samples. The $2\theta$ angle is fixed at 44$^\circ$. The radial axis $\alpha$ of the polar plot goes from 0$^\circ$ (center) to 90$^\circ$ (edge).}
    \label{fig:pole_figure}
\end{figure}

\begin{figure}
    \centering
    \includegraphics[width=10cm]{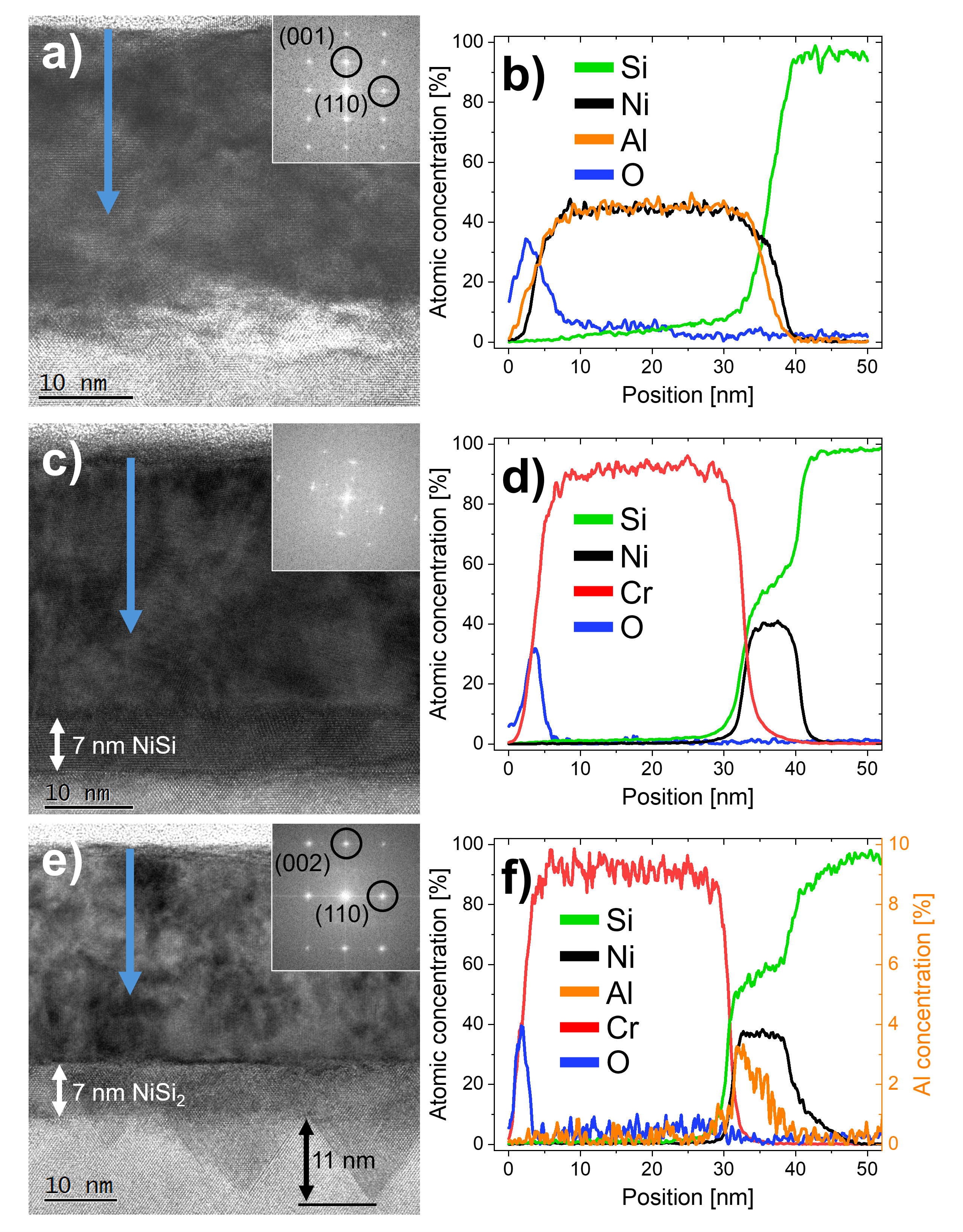}
    \caption{Cross-section bright field-TEM image of sample 1,\textbf{a)}, sample 3, \textbf{c)}, and sample 4, \textbf{e)}. For each TEM image the FFT of the Cr layer is shown in the corner. \textbf{b)}, \textbf{c)}, and \textbf{f)} are atomic concentration profiles measured by EDS. The scan location and direction are indicated on the corresponding TEM by the blue arrow. In f) the vertical scale on the right applies to the Al concentration.}
    \label{fig:TEM_EDS}
\end{figure}

The AFM results are summarized in figure \ref{fig:AFM} where a)-d) are scans covering an area of 2x2\,\textmu m. Samples 1,2, and 4 were smooth and homogeneous with an arithmetic average roughness (Ra) of 0.35, 0.25, and 0.13\,nm, respectively. The surface of sample 3, figure \ref{fig:AFM}c), had the highest roughness with an Ra of 0.51\,nm and had an inhomogeneous topography. The 20x20\,\textmu m-sized scan, \ref{fig:AFM}e), revealed the presence of grooves which had a depth of $\sim2\,\text{nm}$. Presumably, these grooves were formed at the grain boundaries of the Cr layer. Some grains were quite smooth with height variation below 1\,nm while other grains contained many point-like defects that had a lateral size of around 100\,nm and a depth of 2\,nm.
\\
\\
Results from the 2$\theta$-$\theta$ XRD scans are shown in figure \ref{fig:T2T}. Two NiAl peaks were observed in sample 1: NiAl(001) and the higher order NiAl(002), at $2\theta=30.8^\circ$ and 64.1$^\circ$, respectively. This shows that the layer had a (001)-oriented lattice. On the other hand, the Cr layer of sample 2 had a (011) orientation as indicated by the corresponding peak. Such an orientation is explained by
the close-packed Cr(011) plane having the lowest surface energy \cite{fu09}. Sample 3 showed a NiSi peak at $2\theta=32^\circ$ indicating that the Ni layer had undergone a silicidation reaction. Both the Cr(002) and the Cr(011) reflection were measured revealing that the layer was polycrystalline. Regarding sample 4, no NiSi peaks are found. This is a first indication of epitaxial NiSi$_2$ formation with NiSi$_2$(001)//Si(001) and NiSi$_2$(100)//Si(100) \cite{geenen18}. However, cube-on-cube epitaxial NiSi$_2$ could not be identified reliably by a 2$\theta$-$\theta$ scan since its (002) and (004) reflections overlap with Si(002) and Si(004). Regarding the Cr layer of sample 4, only the (002) peak was found, at 64.6$^\circ$, which is consistent with the layer being monocrystalline. However, to make strong conclusion on epitaxial growth, further measurements are required which is subject of the following paragraphs.
\\
\\
In order to further examine the crystallinity of the NiAl and Cr layers, XRD pole figure measurements were done. An XRD pole figure scan makes a map of the orientation distribution of a specific lattice plane distance by measuring diffraction intensity as the sample is tilted and rotated relative to the X-ray beam. The sample was tilted from $\alpha=0^\circ$ to 90$^\circ$ and rotated from $\beta=0^\circ$ to 360$^\circ$. For all samples the $2\theta$ angle was fixed at 44$^\circ$ which probes both the NiAl$\{011\}$ and the Cr$\{011\}$ planes while avoiding the signal from Si$\{022\}$ which peaks at $2\theta=47.3^\circ$. The pole figure of the NiAl sample, figure \ref{fig:pole_figure}a), shows four distinct peaks at $\alpha$=45$^\circ$ coming from $\{011\}$ planes that were angled at $45^\circ$ with respect to the substrate. Four additional peaks were measured at $\alpha$=90$^\circ$ as seen at the edge of the pole figure. These correspond to NiAl$\{011\}$ planes that have a 90$^\circ$ angle with the substrate. The result from the pole figure is consistent with an epitaxial NiAl layer. The pole figure of sample 4, figure \ref{fig:pole_figure}d), also shows four peaks at $\alpha$=45$^\circ$ as well as four peaks at $\alpha$=90$^\circ$. This shows that epitaxial growth of Cr occurred in the same crystal orientation that the NiAl layer had on Si. Sample 3 had four peaks at $\alpha$=45$^\circ$, as shown in figure \ref{fig:pole_figure}c), while no peaks were detected at 90$^\circ$. The distinctive peaks at 45$^\circ$ indicate an epitaxial Cr layer though considering the absence of peaks at 90$^\circ$, it could be tentatively concluded that only a fraction of the Cr grains was epitaxial. Lastly, the pole figure of sample 2, figure \ref{fig:pole_figure}b), showed no peaks and therefore the Cr layer had no epitaxial alignment with the substrate.
\\
\\
TEM and EDS were done for samples 1,3, and 4; the results of which are summarized in figure \ref{fig:TEM_EDS}. The TEM image of the NiAl sample, figure \ref{fig:TEM_EDS}a), confirmed that the layer was monocrystalline. This was made clear by the Fast Fourier Transform (FFT) of the NiAl layer where NiAl(001) and NiAl(110) were identified. The Si/NiAl interface was not sharp indicating that NiAl had diffused into the substrate. This is also seen in the EDS scan, figure \ref{fig:TEM_EDS}b), around position 35\,nm around which gives a strong indication of NiSi$_2$. Ni diffused a few nanometers deeper into the Si substrate than Al which was also observed in previous work \cite{yakushiji19}. The top surface of NiAl was oxidized due to air exposure, as seen by the peak in O concentration. The Al concentration was larger than the Ni concentration at this location leading to the conclusion that the oxide layer was primarily composed of oxidized Al.
\\\\
The TEM of sample 3, figure \ref{fig:TEM_EDS}c), shows the Cr layer to be polycrystalline. Below the Cr layer, NiSi was found to be 7\,nm thick. The expected NiSi thickness formed from 3\,nm thick Ni was estimated by the following calculation: The Ni concentration of fcc Ni is 91 Ni atoms/nm$^3$ while for NiSi this is 41 Ni atoms/nm$^3$. The ratio of these two concentrations is 2.2 and is also the layer thickness ratio $t_{NiSi}/t_{Ni}$. In other words, if all the Ni atoms have reacted to form NiSi, the NiSi layer has to be 2.2 thicker than the initial Ni. For 3\,nm of Ni, this leads to the 6.6\,nm of NiSi which is close the to measured thickness of 7\,nm. The Si/NiSi interface was sharp while around the NiSi/Cr interface significant diffusion of Cr into the NiSi layer was found, as seen in figure \ref{fig:TEM_EDS}d). Additionally, Si was diffused deeper into the Cr layer than Ni was by about 2\,nm.
\\
\\
TEM imaging of sample 4, shown in figure \ref{fig:TEM_EDS}e), confirmed the epitaxial relation of the Cr layer with the Si substrate as measured by the XRD pole figure. This was made clear by the Cr(002) and the Cr(110) reflections identified on the FFT. The NiSi$_2$ layer was 7\,nm thick, excluding the triangular facets. Bulk NiSi$_2$ has a Ni concentration of 25 atoms/nm$^3$ leading to a predicted NiSi$_2$ thickness of 10.7\,nm. The discrepancy with the 7\,nm thickness can be explained by considering the triangular facets which reach as deep as 11\,nm into the Si substrate. The facets were defined by $\{111\}$ planes of Si and NiSi$_2$ which is typical for epitaxial growth of NiSi$_2$ on Si(001) \cite{tung83,falke04,mi09,geenen18}. The presence of these facets could be an indication that the Si(111)/NiSi$_2$(111) interfacial energy had a lower interfacial energy than than Si(001)/NiSi$_2$(001) interface. The EDS scan, shown in figure \ref{fig:TEM_EDS}f), revealed that Ni diffused deeper into the substrate compared to sample 3. The scan direction was chosen such to avoid the faceted NiSi$_2$ region. Thus, any Ni atoms found for positions past 40\,nm are part of the Si phase and not part of the NiSi$_2$ phase. Lastly, the Al layer, which was deposited on the Ni layer, was detected throughout the NiSi$_2$ layer.

\section{Discussion}
The solid phase reaction of 3\,nm Ni with Si resulted in formation of 7\,nm thick NiSi. Previous work has shown that Ni layers below 4\,nm react to form NiSi$_2$ after annealing above 500$^\circ$C \cite{luo10,geenen18}. In another experiment it was found that 3\,nm of Ni forms NiSi starting from 200$^\circ$C and remains stable up until anywhere between 400 and 500$^\circ$C \cite{fouet14}. This is consistent with our work where sample 3 is annealed at 400$^\circ$C and NiSi is formed.
\\\\
The sample with 0.3\,nm Al added, sample 4, resulted in the formation of epitaxial NiSi$_2$. The effect of alloying nickel silicides with Al has been investigated before and was found to reduce the temperature at which NiSi$_2$ is formed \cite{allenstein05}. This same effect could explain the formation of NiSi$_2$ during NiAl growth at 400$^\circ$C. As seen in the EDX scan of the NiAl sample (figure \ref{fig:TEM_EDS}b)) Al has diffused together with Ni into the Si forming epitaxial NiSi$_2$.
\\\\
Two mechanisms have been proposed to explain why Al promotes NiSi$_2$ formation over NiSi. One mechanism is related to the mixing entropy created when Al --- or another alloying element --- has mixed with the silicide \cite{geenen18,luo10,detavernier00}. Since Al is more soluble in NiSi$_2$ than in NiSi, a higher mixing entropy is created in the former. The second proposed mechanism by which Al can promote NiSi$_2$ growth is related to the increased lattice parameter of NiAl$_{x}$Si$_{2-x}$ compared to NiSi$_2$. The interfacial energy of epitaxial Si/NiAl$_{x}$Si$_{2-x}$ is a function of the lattice mismatch and should be minimal when there is zero lattice mismatch. Without Al, NiSi$_2$ has a 0.6\% smaller lattice compared to Si at room temperature. At 480$^\circ$C the lattice mismatch is reduced to zero because NiSi$_2$ has a larger thermal expansion coefficient compared to Si \cite{smeets09}. Adding Al to NiSi$_2$ expands the lattice by replacing Si atoms \cite{allenstein05,richter03,raghavan05}. Work by \textit{Richter et al.} found that the lattice parameter of NiAl$_{x}$Si$_{2-x}$ changes roughly linearly with the Al concentration and is given by $a_{NiAl_{x}Si_{2-x}}=a_{NiSi_2}+0.097x$ (in \AA) \cite{richter03}. Inside the NiSi$_2$ layer of sample 4, the Al atomic concentration is around 2\%, as shown by EDS in figure \ref{fig:TEM_EDS}f). This translates to $x=0.06$ and an increase of the lattice parameter of 0.006\,\AA. This results in perfect lattice matching with Si at already $\sim$400$^\circ$C compared to 480$^\circ$C for NiSi$_2$.
\\\\
The Cr layer with NiSi$_2$ as the bridging layer was fully epitaxial. This was shown both by the pole figure and the TEM image, figure \ref{fig:pole_figure}d) and figure \ref{fig:TEM_EDS}e), respectively. From this it can be concluded that NiSi$_2$ enables epitaxial growth of both NiAl and Cr. Regarding the NiSi/Cr sample, it was shown by the 2$\theta$-$\theta$ scan and by TEM that the Cr was polycrystalline. However, the pole figure did indicate that some degree of epitaxy was present in the layer, see figure \ref{fig:pole_figure}c). The likely explanation therefore is that the Cr grains with (002) orientation were epitaxially grown while grains with (111) orientation were polycrystalline. This means that NiSi was acting as a seed layer for the Cr layer such that only a fraction of the Cr grains were epitaxial. Previous work by \textit{Detavernier et al.} discovered that NiSi grown on Si(001) has a complex grain texture which includes multiple axiotaxy components (i.e., off-axis fiber texture) and one epitaxial relation \cite{detavernier03}. The epitaxial relation is described as NiSi(014)//Si(001) and NiSi(200) nearly parallel with Si(010). Assuming a similar NiSi grain texture for the NiSi layer in sample 3, it can be deduced that epitaxial Cr grains were templated by epitaxial NiSi grains. The non-epitaxial Cr grains were not in contact with epitaxial NiSi grains and would therefore be polycrystalline.

\section{Conclusion}
In this work, epitaxial growth of NiAl on Si(001) was investigated. During the NiAl deposition the layer reacted with Si to form epitaxial NiSi$_2$. From this observation, the hypothesis was made that NiSi$_2$ can be used as a seed layer to epitaxially grow metal layers on Si(001). To test this hypothesis, Cr was deposited on NiSi and NiSi$_2$. The NiSi$_2$/Cr bilayer was found to result in epitaxial Cr with the same lattice orientation as NiAl. From this, it was concluded that NiSi$_2$ was not merely a byproduct during the NiAl growth but instead it is essential for epitaxial growth. Moreover the NiAl can be swapped for other bcc metals with a similar lattice parameter e.g., Cr. The NiSi/Cr sample resulted in a polycrystalline Cr layer. Utilizing NiSi$_2$ as a seed layer is expected to enable the epitaxial growth of other metals and on Si wafers. This makes implementation of epitaxial metals more feasible to integrate for industrial chip fabrication.

\section{Acknowledgments}
We acknowledge funding from the Research Foundation – Flanders (FWO) through grant 1S10125N. This work was also supported by the FWO WEAVE program G0D7723N and by the KU Leuven BOF funding C14/24/110. Lastly, the authors acknowledge the support of imec's fab and hardware team as well as Paola Favia for TEM and EDS measurements. The authors declare no conflicts of interest.

\section{References}

\end{document}